\documentclass[12pt]{article}
\begin{document}

\title{Neutrino Mass Matrix with Approximate Flavor Symmetry}
\author{Riazuddin \\
National Centre for Physics, \\
Quaid-i-Azam University,\\
Islamabad, Pakistan.}
\maketitle

\begin{abstract}
Phenomenological implications of neutrino oscillations implied by recent
experimental data on pattern of neutrino mass matrix are disscussed. It is
shown that it is possible to have a neutrino mass matrix which shows
approximate flavor symmetry; the neutrino mass differences arise from flavor
violation in off-diagonal Yukawa couplings. Two modest extensions of the
standard model, which can embed the resulting neutrino mass matix have also
been discussed.
\end{abstract}

The data \cite{1,2,3,4,5} from solar and atmospheric neutrino experiments
provide evidence for neutrino mass and mixing. As such lepton flavor is
violated providing evidence for physics beyond the standard model (SM). The
present experimental data can be explained by neutrino oscillations with
mass squared differences and mixing angles having following best fit values 
\cite{6} 
\begin{eqnarray}
\Delta m_{{atm}}^2 &\equiv &\Delta m_{23}^2=\left( 2.7\pm 0.4\right) \times
10^{-3}{\ eV}^2  \nonumber \\
\sin ^2\theta _{23} &\equiv &\sin ^22\theta _1=1.00\pm 0.04  \label{01}
\end{eqnarray}
\begin{eqnarray}
\Delta m_{{solar}}^2 &\equiv &\Delta m_{12}^2=\left( 7.1\pm 0.6\right)
\times 10^{-5}{\ eV}^2  \nonumber \\
\tan ^2\theta _{12} &\equiv &\tan ^2\theta _3=0.45\pm 0.06  \label{02}
\end{eqnarray}
Further the CHOOZ \cite{7} experiment gives a bound 
\begin{equation}
\left| U_{e3}\right| ^2\equiv \sin ^2\theta _2<4\times 10^{-2}  \label{03}
\end{equation}
The purpose of this paper is to interpret these results in terms of small
off-diagonal perturbations of a degenerate diagonal mass matrix in flavor
basis for light Majorana neutrinos. In this approach there is no fundamental
distinction between masses of neutrinos of different flavor; the mass
differences arise from small flavor violation of off diagonal Yukawa
coupling constants. Further it is shown that the neutrino mass differences
do not in anyway constraint the absolute value of neutrino mass. The
constraint on it will come from neutrinoless double-beta decay experiments,
cosmology and direct laboratory experiments, e.g. tritium beta-decay.

Lastly we will try to embed the resulting neutrino mass matrix in two
different modest extensions of the standard model. In one version the
minimal standard model of particle interactions is extended to include a
Higgs triplet to introduce Majorana mass. In the other the standard
electroweak gauge group is extended to $SU_L\left( 2\right) \times U_e\left(
1\right) \times U_\mu \left( 1\right) \times U_\tau \left( 1\right) $
leading to off-diagonal mass matrix for light (Majorana) neutrinos which can
act as a small perturbation to a degenerate diagonal matrix.The implications
of these models on neutrinoless double beta decay are also discussed.

Let us consider a Majorana mass matrix in ($e,\,\mu ,\,\tau $) basis 
\begin{equation}
M_\nu =m_0\left( 
\begin{array}{ccc}
a_{ee} & a_{e\mu } & a_{e\tau } \\ 
a_{e\mu } & a_{\mu \mu } & a_{\mu \tau } \\ 
a_{e\tau } & a_{\mu \tau } & a_{\tau \tau }
\end{array}
\right)   \label{04}
\end{equation}
It is convenient to define the neutrino mixing angles as follows 
\begin{equation}
\left( 
\begin{array}{c}
\nu _e \\ 
\nu _\mu  \\ 
\nu _\tau 
\end{array}
\right) =U\left( 
\begin{array}{c}
\nu _1 \\ 
\nu _2 \\ 
\nu _3
\end{array}
\right)   \label{05}
\end{equation}
where 
\begin{equation}
\left( 
\begin{array}{ccc}
c_2c_3 & c_2s_3 & s_2e^{i\delta } \\ 
-c_1s_3-s_1s_2c_3e^{i\delta } & c_1c_3-s_1s_2s_3e^{i\delta } & s_1c_2 \\ 
s_1s_3-c_1s_2c_3e^{i\delta } & -s_1c_3-c_1s_2s_3e^{i\delta } & c_1c_2
\end{array}
\right)   \label{06}
\end{equation}
with $c_i=\cos \theta _i$ and $s_i=\sin \theta _i$. We shall put $\delta $
as well as Majorana phases to be zero. In view of Eqs. (\ref{01}), (\ref{02}%
) and (\ref{03}), we shall take 
\[
s_2=0,\,c_2=1.
\]
The diagonalization of mass matrix (\ref{04}) give the relations 
\begin{eqnarray}
m_1c_3^2+m_2s_3^2 &=&a_{ee}  \nonumber \\
m_1c_1^2s_3^2+m_2c_1^2c_3^2+m_3s_1^2 &=&a_{\mu \mu }  \nonumber \\
m_1s_1^2s_3^2+m_2s_1^2c_3^2+m_3c_1^2 &=&a_{\tau \tau }  \nonumber \\
c_1s_3c_3(-m_1+m_2) &=&a_{e\mu }  \nonumber \\
s_1c_3s_3(m_1-m_2) &=&a_{e\tau }  \nonumber \\
-c_1s_1\left( m_1s_3^2+m_2c_3^2-m_3\right)  &=&a_{\mu \tau }  \label{07}
\end{eqnarray}
Taking now $c_1=\frac 1{\sqrt{2}}$, $s_1=\mp \frac 1{\sqrt{2}}$, we obtain 
\begin{eqnarray}
a_{e\mu } &=&\pm a_{e\tau }=\frac 1{\sqrt{2}}s_3c_3\left( -m_1+m_2\right)  
\nonumber \\
a_{\mu \tau } &=&\pm \frac 12\left[ \left( m_1s_3^2+m_2c_3^2\right)
-m_3\right]   \nonumber \\
a_{\mu \mu } &=&a_{\tau \tau }=\frac 12\left( m_1s_3^2+m_2c_3^2+m_3\right)  
\nonumber \\
a_{ee} &=&m_1c_3^2+m_2s_3^2  \label{08}
\end{eqnarray}
Further defining 
\[
\Delta m_{12}^2=m_2^2-m_1^2{,}\Delta m_{23}^2=m_3^2-m_2^2{,}\Delta
m_{13}^2=m_3^2-m_1^2
\]
so that 
\[
\Delta m_{12}^2+\Delta m_{23}^2=\Delta m_{13}^2
\]
and in view of $\Delta m_{12}^2\ll \Delta m_{23}^2$ as indicated in Eqs. (%
\ref{01}), (\ref{02}), we can take 
\[
m_1\simeq \pm m_2
\]
Thus we have two possibilities for mass matrix $m_1=m_2=m_0;\,m_1=-m_2=m_0:$%
\begin{eqnarray*}
M_\nu  &=&m_0\left( 
\begin{array}{ccc}
1 & 0 & 0 \\ 
0 & \frac 12\left( 1+a\right)  & \pm \frac 12\left( 1-a\right)  \\ 
0 & \pm \frac 12\left( 1-a\right)  & \frac 12\left( 1+a\right) 
\end{array}
\right)  \\
M_\nu  &=&m_0\left( 
\begin{array}{ccc}
\cos 2\theta _3 & -\frac 1{\sqrt{2}}\sin 2\theta _3 & \mp \frac 1{\sqrt{2}%
}\sin 2\theta _3 \\ 
-\frac 1{\sqrt{2}}\sin 2\theta _3 & \frac 12\left( \cos 2\theta _3+a\right) 
& \pm \frac 12\left( \cos 2\theta _3-a\right)  \\ 
\mp \frac 1{\sqrt{2}}\sin 2\theta _3 & \pm \frac 12\left( \cos 2\theta
_3-a\right)  & \frac 12\left( \cos 2\theta _3+a\right) 
\end{array}
\right) 
\end{eqnarray*}
where $a=m_3/m_0$. If we do not want to commit to any particular value of $%
\theta _3$, then we have the first case with the following subcases
corresponding to $m_0=0$, $a=0$, $a=-1,\,1,\,-2,\,2$ respectively 
\begin{eqnarray*}
i) &:&\frac{m_3}2\left( 
\begin{array}{ccc}
0 & 0 & 0 \\ 
0 & 1 & 1 \\ 
0 & 1 & 1
\end{array}
\right) \,\,\,\,\,\,\,\,\,\,\,\,\,\,\cite{8} \\
\,ii) &:&m_0\left( 
\begin{array}{ccc}
1 & 0 & 0 \\ 
0 & 0 & 1 \\ 
0 & 1 & 0
\end{array}
\right) \,\,\,\,\,\,\,\,\,\,\,\,\,\,\,\cite{8} \\
\,iii) &:&m_0\left( 
\begin{array}{ccc}
1 & 0 & 0 \\ 
0 & 1 & 0 \\ 
0 & 0 & 1
\end{array}
\right)  \\
\,\,iv) &:&\frac{m_0}2\left( 
\begin{array}{ccc}
2 & 0 & 0 \\ 
0 & -1 & 3 \\ 
0 & 3 & -1
\end{array}
\right) \,\,\,\,\,\,\,\cite{9} \\
v) &:&\frac{m_0}2\left( 
\begin{array}{ccc}
2 & 0 & 0 \\ 
0 & 3 & -1 \\ 
0 & -1 & 3
\end{array}
\right) 
\end{eqnarray*}

In order to generate $\Delta m_{12}^2$ and $\Delta m_{23}^2$, we will now
concentrate on choice (iii) which preserves flavor and add to it a small
perturbation which violates flavor in off-diagonal matrix elements: 
\begin{equation}
M_\nu =m_0\left( 
\begin{array}{ccc}
1 & \varepsilon _{12} & \varepsilon _{13} \\ 
\varepsilon _{12} & 1 & \varepsilon _{23} \\ 
\varepsilon _{13} & \varepsilon _{23} & 1
\end{array}
\right)  \label{09}
\end{equation}
where $\varepsilon _{ij}\ll 1$. The diagonalization gives 
\begin{equation}
m_i=m_0\left( 1-x_i\right)  \label{10}
\end{equation}
where $x_i$ ($i=1,\,2,\,3$) are roots of cubic equation 
\begin{equation}
x^3-\left( \varepsilon _{12}^2+\varepsilon _{13}^2+\varepsilon
_{33}^2\right) x+2\varepsilon _{12}\varepsilon _{13}\varepsilon _{23}=0
\label{11}
\end{equation}
The choice $\varepsilon _{12}=\varepsilon _{13}=\varepsilon
_{23}=\varepsilon $ will give the roots $\left( \varepsilon ,\,\varepsilon
,\,-2\varepsilon \right) $ and thus will not lift the degeneracy between $%
m_1 $ and $m_2$. To lift this degeneracy we take $\varepsilon
_{12}=\varepsilon _{13}=\varepsilon +\delta $, $\varepsilon
_{23}=\varepsilon $ with $\delta /\varepsilon \ll 1$. Then the roots to the
first order in $\delta /\varepsilon $ are $\varepsilon $, $\varepsilon
\left( 1+\frac 43\delta /\varepsilon \right) $, -2$\varepsilon \left(
1+\frac 23\delta /\varepsilon \right) $ so that 
\begin{eqnarray}
m_1 &=&m_0\left[ 1-\varepsilon -\frac 43\delta \right]  \nonumber \\
m_2 &=&m_0\left[ 1-\varepsilon \right]  \nonumber \\
m_3 &=&m_0\left[ 1+2\varepsilon +\frac 43\delta \right]  \label{12}
\end{eqnarray}
\begin{eqnarray}
\Delta m_{12}^2 &\approx &\frac 83m_0^2\delta \left( 1-\varepsilon \right)
\simeq \frac 83m_0^2\delta  \nonumber \\
\Delta m_{32}^2 &\approx &6\varepsilon m_0^2  \label{13}
\end{eqnarray}
This gives 
\begin{equation}
\frac \delta \varepsilon =\frac 94\frac{\Delta m_{12}^2}{\Delta m_{32}^2}%
\simeq 5.9\times 10^{-2}  \label{14}
\end{equation}
\[
\sqrt{\varepsilon }m_0\simeq 2.1\times 10^{-2}{eV} 
\]
so that $m_0$ is not constrained. However, $m_0$ is constrained by WMAP data 
\cite{10},$3m_0<0.71$ eV. When analyzed in conjunction with neutrino
oscillation, it is found that mass eigenvalues are essentially degenerate
with $3m_0>0.4$ eV \cite{11} The above limits put limits on $\varepsilon $ : 
$7.9\times 10^{-3}<\varepsilon <2.5\times 10^{-2}$.

It is pertinent to remark that in view of small mass differences involved
[cf. Eqs. (1) and (2)], and that analysis of WMAP data in conjunction with
neutrino oscillation imply essentially degenrate mass eigenvalues, it is
quite natural to remove the degenracy by small perturbations. Such a
procedure should be stable as the parameter $\varepsilon $ turns out to be
small in view of limit on it given above. The $\delta /\varepsilon \simeq
6\times 10^{-2}$ may indicate fine tuning but this a problem common to
Yukawa couplings even in charged sector, the solution of which has so far
evaded us.

Next we consider an extension of the standard electroweak model to 
\[
G\equiv SU_L\left( 2\right) \times U_e\left( 1\right) \times U_\mu \left(
1\right) \times U_\tau \left( 1\right) 
\]
which naturally gives an off diagonal mass matrix for light neutrinos \cite
{12}. In addition to usual fermions $L_i=\left( 
\begin{array}{c}
\nu ^i \\ 
e^i
\end{array}
\right) _L,\,e_R^i$, there are three right handed $SU_L\left( 2\right) $%
-singlet neutrinos $N_R^i$ which carry $U_i\left( 1\right) $ quantum numbers 
$\left( -1,\,1,\,0\right) $, $\left( 1,\,0,\,-1\right) $ and $\left(
0,\,-1,\,1\right) $. Further in addition to $SU_L\left( 2\right) $ Higgs
doublets, there are three Higgs $SU_L\left( 2\right) $ singlets $\Sigma ^i$
with $U_i\left( 1\right) $ quantum numbers $\left( 1,\,-1,\,0\right) $, $%
\left( 1,\,0,\,-1\right) $ and $\left( 0,\,1,\,-1\right) $. In order to
introduce diagonal terms we also introduce a right handed neutrino $N$ and a
corresponding Higgs boson $\Sigma $ which are $SU_L\left( 2\right) $ and $%
U_i\left( 1\right) $ singlets. The symmetry is spontaneously broken by
giving vacuum expectation values to Higgs boson $\phi ^{\left( i\right) }$, $%
\Sigma ^{\left( i\right) }$ and $\Sigma :$%
\begin{equation}
\left\langle \phi ^{\left( i\right) }\right\rangle =\frac{v_i}{\sqrt{2}}{,}%
\left\langle \Sigma ^{\left( i\right) }\right\rangle =\frac{\Lambda _i}{%
\sqrt{2}}{,}\left\langle \Sigma \right\rangle =\frac \Lambda {\sqrt{2}}
\label{15}
\end{equation}
where for simplicity we shall take $v_1=v_2=v_3=v$ and $\Lambda _1=\Lambda
_2=\Lambda _3=\Lambda $ (any difference can be absorbed in the corresponding
Yukawa couplings of fermions with Higgs bosons). The resulting mass matrix
through seesaw mechanism for light neutrinos is 
\begin{equation}
M_\nu =\frac{v^2}{2M_R}\left( 
\begin{array}{ccc}
h_1^{\left( 1\right) }h_1^{\left( 1\right) } & h_1^{\left( 2\right)
}h_2^{\left( 3\right) } & h_1^{\left( 2\right) }h_3^{\left( 1\right) } \\ 
h_1^{\left( 2\right) }h_2^{\left( 3\right) } & h_2^{\left( 2\right)
}h_2^{\left( 2\right) } & h_2^{\left( 3\right) }h_3^{\left( 1\right) } \\ 
h_1^{\left( 2\right) }h_3^{\left( 1\right) } & h_2^{\left( 3\right)
}h_3^{\left( 1\right) } & h_3^{\left( 3\right) }h_3^{\left( 3\right) }
\end{array}
\right)   \label{16}
\end{equation}
where $M_R=\frac{f\Lambda }{\sqrt{2}}.$ We now put $h_i^{\left( i\right) }=h$%
, $\frac{h_1^{\left( 2\right) }h_2^{\left( 3\right) }}{h^2}=\varepsilon _{12}
$, $\frac{h_1^{\left( 2\right) }h_3^{\left( 1\right) }}{h^2}=\varepsilon
_{13}$, $\frac{h_2^{\left( 3\right) }h_3^{\left( 1\right) }}{h^2}%
=\varepsilon _{23}$. This gives the desired matrix (\ref{09}) with $m_0=%
\frac{h^2v^2}{2M_R}$. If $h$ is of the order of $1$ and $m_0$ is few
electron volts, then one requires $M_R\simeq 10^{13}$ GeV.

Another simple way to obtain the neutrino mass matrix (\ref{09}) is to
include a Higgs triplet $\Delta \left( 3,2\right) :$ $\left( \Delta
^{++},\,\Delta ^{+},\,\Delta ^0\right) $ with the interaction\cite{13,14}

\begin{equation}
\mathcal{L}=h_{ij}L_i^Ti\tau _2\left( \vec{\tau}.\vec{\Delta}\right) L_j
\label{e17}
\end{equation}
generating neutrino mass matrix 
\begin{equation}
\left( M_\nu \right) _{ij}=2h_{ij}\left\langle \Delta ^0\right\rangle 
\label{e18}
\end{equation}
Taking $h_{11}=h_{22}=h_{33}=h$ and $\frac{h_{12}}h=\varepsilon _{12}$, $%
\frac{h_{13}}h=\varepsilon _{13}$, and $\frac{h_{23}}h=\varepsilon _{23};$ $%
m_0=2h\left\langle \Delta ^0\right\rangle $ we get the desired mass matrix.
If $h$ is of order unity, and $m_0$ is a few electron volts, then $%
\left\langle \Delta ^0\right\rangle $ is of order of eV, which appears to be
unnatural. It is well known that the coupling of $\Delta $ to gauge field
will alter the $\rho $ parameter $\left( \rho \equiv \frac{m_w^2}{m_z^2\cos
^2\theta _w}\right) $ by 
\begin{equation}
\Delta \rho \equiv \left( 1-\rho \right) =4\left\langle \Delta
^0\right\rangle ^2/v^2  \label{e19}
\end{equation}
The present limit on $\Delta \rho $ $\left( \leq 0.02\right) $ only requires 
$\left\langle \Delta ^0\right\rangle <25$ GeV. To give a more natural look
to small value of $\Delta ^0$, lepton flavor violating term of the form $\mu
\left( \phi ^{\dagger }\Delta \tilde{\phi}\right) +H.C$ is added to $L$
conserving Higgs potential. Thus \cite{14} 
\begin{equation}
\left\langle \Delta ^0\right\rangle =\frac{\mu v^2}{2M_\Delta ^2}\simeq
3\times 10^4\mu \left( \frac{{GeV}^2}{M_\Delta ^2}\right)   \label{e20}
\end{equation}
For $\left\langle \Delta ^0\right\rangle \sim $eV $\mu $ has to be quite
small $\left( \simeq 3{eV}\right) $ if $M_\Delta $ is $\sim 1$TeV. If $\mu
\sim 1$GeV$,M_\Delta \simeq 3\times 10^6$GeV.

Finally we discuss the implications of above models in neutrinoless double
beta decay. In the first model at tree level there is only the standard
contribution through light Majorana $\nu _e$. In the second model, there
could be additional contribution at tree level through diagram in figure 1.

This give rise to an effect $\Delta L=2$ interaction which can be written as 
\cite{15n} 
\begin{eqnarray}
H_{\Delta L=2} &=&-G_F^2\epsilon _1^{ee}\bar{u}\left( 1+\gamma _5\right) d%
\bar{u}\gamma _\mu \left( 1-\gamma _5\right) d  \nonumber \\
&&\,\,\,\,\,\,\,\,\,\,\,\,\,\,\,\,\,\,\,\,\,\,\,\,\bar{e}\gamma ^\mu \left(
1-\gamma _5\right) \frac 1{\gamma -q}C\bar{e}^T  \label{21n}
\end{eqnarray}
where [$h_{11}=h$] 
\begin{equation}
\epsilon _1^{ee}=\frac{h_u\left( h\mu \frac v{\sqrt{2}}\right) }{u\sqrt{2}%
G_Fm_\Delta ^2m_H^2}  \label{22n}
\end{equation}
On using Eq. (\ref{e20}) and that $m_0=2h\left( \Delta ^0\right) $, we
obtain 
\[
\epsilon _1^{ee}=\frac{\left( h_u\right) m_0}{8\sqrt{2}G_Fm_H^2\frac v{\sqrt{%
2}}}
\]
With $\frac v{\sqrt{2}}\simeq 175$ GeV, we obtain $\epsilon
_1^{ee}=h_u\left( \frac{m_0}{eV}\right) \left( 5\times 10^{-10}\right) $.
One may expect \cite{15n} $h_u=m_u/m_W\simeq 5\times 10^{-5}$ so that $%
\epsilon _1^{ee}$ is much too small compared to the upper limit on $\epsilon
_1^{ee}\left( \leq 10^{-8}\right) $ found in \cite{15n} from the neutrino
mass contribution to $\beta \beta _0\nu $ ($m_\nu <1$ eV).

To sum up we have shown that it is possible to interpret the neutrino mass
differences and mixings with approximate flavor symmetry. The small flavor
violation, manifested in neutrino mass differences, is in the Yukawa
coupling constants $h_{ij}$ which determine the off-diagonal matrix elements
with $\varepsilon _{ij}\left( i\neq j\right) \equiv \frac{h_{ij}^2\left(
i\neq j\right) }{h^2}\sim 10^{-3}$. This may have some phenomenological
implications. We have also discussed two modest extensions of the standard
model which can embed the resulting neutrino mass matrix.

{\bf Note added}: It has been called to my attention by A.Zee that the
matrix(9) in this paper is similar to $\widetilde{M}_0$ with $y=-1$, given
by X.G.He and A.Zee [16]. Our motivation is rather different.

{\bf Acknowledgement}

This work was supported by the grant from Pakistan Council of Science and
Technology.

{\bf Figure Caption:} 

Figure 1: The vector-scalar exchange diagram for $\beta \beta_{0\nu}$

\end{document}